\documentclass[10pt]{article}
\usepackage[a4paper,right=1in,left=1in,top=1in,bottom=1in]{geometry}
\pdfoutput=1

\usepackage{amsmath,amssymb}

\usepackage{changepage}

\usepackage[utf8x]{inputenc}

\usepackage{textcomp,marvosym}

\usepackage{cite}

\usepackage{nameref,hyperref}

\usepackage[right]{lineno}

\usepackage{microtype}
\DisableLigatures[f]{encoding = *, family = * }

\usepackage[table]{xcolor}

\usepackage{multirow}
\usepackage[export]{adjustbox}

\usepackage{array}

\newcolumntype{+}{!{\vrule width 2pt}}

\newlength\savedwidth

\newcommand\thickhline{\noalign{\global\savedwidth\arrayrulewidth\global\arrayrulewidth 2pt}%
\hline
\noalign{\global\arrayrulewidth\savedwidth}}

\usepackage[aboveskip=1pt,labelfont=bf,labelsep=period,justification=raggedright,singlelinecheck=off]{caption}

\makeatletter
\renewcommand{\@biblabel}[1]{\quad#1.}
\makeatother

\begin{document}
\vspace*{0.2in}

\begin{center}
{\Large
\textbf\newline{Impact of COVID-19 Policies and Misinformation on Social Unrest} 
}
\newline

Martha Barnard\textsuperscript{1},
Radhika Iyer \textsuperscript{1,2},
Sara Y Del Valle \textsuperscript{1},
Ashlynn R Daughton \textsuperscript{1*}
\\
\bigskip
\textbf{1} A-1 Information Systems and Modeling, Los Alamos National Lab, Los Alamos, NM, USA
\\
\textbf{2} Department of Political Science and Department of Computing, Data Science, and Society, University of California, Berkeley, Berkeley, CA, USA
\\
* adaughton@lanl.gov

\end{center}
\section*{Abstract}

The novel coronavirus disease (COVID-19) pandemic has impacted every corner of earth, disrupting governments and leading to socioeconomic instability. This crisis has prompted questions surrounding how different sectors of society interact and influence each other during times of change and stress. Given the unprecedented economic and societal impacts of this pandemic, many new data sources have become available, allowing us to quantitatively explore these associations. Understanding these relationships can help us better prepare for future disasters and mitigate the impacts.  Here, we focus on the interplay between social unrest (protests), health outcomes, public health orders, and misinformation in eight countries of Western Europe and four regions of the United States. We created $1-3$ week forecasts of both a binary protest metric for identifying times of high protest activity and the overall protest counts over time. We found that for all regions, except Belgium, at least one feature from our various data streams was predictive of protests. However, the accuracy of the protest forecasts varied by country, that is, for roughly half of the countries analyzed, our forecasts outperform a naïve model. These mixed results demonstrate the potential of diverse data streams to predict a topic as volatile as protests as well as the difficulties of predicting a situation that is as rapidly evolving as a pandemic.

\section*{Introduction}
The first case of the novel coronavirus disease (COVID-19) was first identified in Wuhan, China in late December 2019. As of October 2021, over 230 million cases have been reported and nearly 5 million people have died ~\cite{WHO2021}. Since early 2020, COVID-19 has dominated discourse about medicine, education, politics, public health, social unrest, and more. Initially, information about COVID-19 and its potential impact was limited and the future path of the pandemic was unclear. However, as the situation evolved, many government organizations, industries, and health departments started to collect and publish data on a daily basis. As such, the pandemic and associated impacts have been closely documented through an unprecedented data revolution  that include all facets of life from healthcare, economic, and other societal impacts. The world economy has been disrupted, with global trading down 13\% - 32\% in 2020~\cite{Ozili2020}. COVID-19 lockdowns caused the closing of schools across the world affecting 1.5 billion students, an increase in domestic violence, and worsening mental health conditions \cite{UNESCO2020, Roesch2020}. A recent study showed that the majority of survey respondents reported having higher levels of anxiety, stress, and panic after the start of the COVID-19 pandemic~\cite{Hussain2020}. There has also been a substantial amount of misinformation surrounding COVID-19 that has spread quickly through social media. One study found that 24\% of tweets associated with COVID-19 were misinformation, and a survey found that 47\% people thought they could not distinguish between reliable and unreliable COVID-19 information~\cite{Kouzy2020, Hussain2020}.  Understanding how a healthcare crisis impacts other aspects of life can help us address future pandemics and other emerging threats. For the purpose of our research, we will focus on the relationship between COVID-19 and protests.

\subsection*{Prior Work}
Now, over a year into the pandemic, the impact of COVID-19 on protests has been studied through a variety of lenses. Some studies have focused on the impact that  non-COVID related protests had on COVID-19 outcomes and found that in some locations, protests increased COVID-19 cases, but impact was generally minimal~\cite{Valentine2020, Dave2020}. Other studies have explored the impact of COVID-19 on protests as a whole. Overall, the total number of conflicts, including peaceful and violent protests and demonstrations, declined during the beginning of the COVID-19 pandemic~\cite{Bloem2021, Berman2020}. However, this trend was geographically heterogeneous and researchers suggested that the relationship between lockdown policies, socioeconomic factors, and the number of protests is more complex. For example, a study of social unrest in the United States published in 2020 found a link between economic inequality, protests, and COVID-19 policy measures. Specifically, they showed that locations with high economic inequality and social distancing restrictions had the most protests~\cite{Iacoella2021}.

Previous studies have found it difficult to determine the impacts of public health orders on human behavior; researchers have shown that individuals often change their behavior in response to their own perceived risk of the disease, rather than reacting to public health orders~\cite{Poletti2012}. Retrospective research on the 1918 influenza pandemic found that public health orders and restrictions had some beneficial impact on disease outcomes, but individuals often changed their behavior based on disease mortality trends within their own communities~\cite{Bootsma2007}. Other studies have focused predominantly on how circumstances other than public health orders alter human behavior. Studies of behavior change during the H1N1 influenza pandemic found that obtaining reliable disease information and higher levels of disease worry were related to participating in mitigation behaviors and vaccination~\cite{bish2011, Liao2011}. In addition, social pressure and past vaccine experience have been found to increase vaccination uptake~\cite{bish2011, Liao2011}. Finally, one meta-analysis found demographic differences in mitigation behavior practices: women were $50\%$ more likely than men to uptake a non-pharmaceutical behaviors, while men were slightly more likely to adopt pharmaceutical changes~\cite{Moran2016}. Thus, despite the apparent disconnect between public health orders and health-conscious behavior, analyzing public health order strength may offer more insight into COVID-19-related protests since many protests documented during the pandemic have been in response to certain health policies such as lockdowns, masks, and vaccine mandates.

In contrast to public health orders, there is substantial previous research exploring the impact of misinformation and the role of social media on protests. The majority of studies on health-related misinformation have focused predominantly on infectious disease and vaccines~\cite{Wang2019}. Wang et al. finds that misinformation surrounding infectious diseases and vaccines tends to be more popular than evidence-based information, even when it is easily available~\cite{Wang2019}. Other researchers have explored the process of misinformation spread and susceptibility to health related misinformation. Roozenbeek et al. assert that there are indicators that can be used to predict both susceptibility to misinformation and resiliency against misinformation, including finding misinformation via social media, age, trust in scientists, and other factors~\cite{Roozenbeek2020}. The authors find that information about COVID-19 on social media is linked to higher susceptibility to misinformation and, notably, Roozenbeek et al. finds links between being susceptible to misinformation and “vaccine hesitancy and a reduced likelihood of complying with public health guidance”~\cite{Roozenbeek2020}. While there are multiple theories about the process of misinformation spread and uptake, there are few studies that directly explore the relationship between misinformation and protests. However, some studies find that misinformation spread can have a bidirectional relationship with protests; some protests may be encouraged by the spread of misinformation, while the protests themselves may instigate spread of misinformation through social media~\cite{ Khan2019, Zervopoulos2020}.  
 
Furthermore, studies have examined the link between social media and protests. Researchers have found that social media has the capacity to draw attention to and mobilize protests~\cite{Karduni1010, Boulianne2020}. Different forms of media vary in their success at mobilization;  traditional TV media has almost no capacity for mobilization, while Twitter is more consistent at mobilizing protests than Facebook~\cite{Boulianne2020}. Korolov et al. built on early efforts of prediction using social media data by classifying messages as stages of mobilization (i.e., sympathy, awareness, motivation, and ability) in order to help determine the timing of an upcoming protest~\cite{Korolov2016}. They found that they could identify Twitter posts indicating mobilization by clustering messages based on the language used, and these messages could subsequently be used to find the probability of a resulting protests~\cite{Korolov2016}. Based on their models, the authors found that it could be possible to use these analyses to predict protests but additional work is needed to extract features from social media that can identify mobilization~\cite{Korolov2016}. Other studies have also performed exploratory analyses on the prediction potential of social media data, finding that both Facebook and Twitter data have bidirectional forecasting potential with protest data, however, this potential exists only for certain protests~\cite{Bastos2015}. In addition, one study of Russian social media found that increased social media activity was related to an increased probability of a protest occurring and protest size~\cite{Enikolopov2020}.

Other researchers have worked on forecasting protests with a variety of data, including social media. Studies have explored how specific protest events, such as the 2016 election protests or the Baltimore protests after the death of Freddie Gray, can be accurately predicted~\cite{Korolov2016, Bahrami2018}. Often, studies implement natural language processing techniques such as sentiment analysis and topic modeling to extract the purpose and/or motivation from social media posts, blogs, or previous event descriptions~\cite{Korolov2016, Kallus2014}. Others used social media data sources such as Google Trends, which provide a normalized search volume for search terms~\cite{Timoneda2021}. While many studies have used solely social media data as a predictor of protests, some have combined a variety of data sources (traditional political and economic data, social media, blogs, etc.)~\cite{Korkmaz2016}. Overall, studies have found that social media data can predict whether a protest will occur, or whether a day will have a significantly higher number of protests than average. However, the majority of studies explore a binary protest outcome, rather than predicting the number of protests over time. In addition, most studies focus on predicting all kinds protests events over a large time frame, or a localized and short-term series of protests.

In this study, we explored how COVID-19 health outcomes, public health orders, and misinformation (measured using Google Trends Data) contributed to COVID-19-related protests and whether these data could accurately forecast the protests. Examining this relationship provides insights into how large, sustained societal changes (such as those brought about by COVID-19), interact with government policy and estimate future human behavior. In addition, the subject of COVID-19-related protests is unique in that even though the protests are global and span a significant amount of time, they center around a specific topic. Both the spatial heterogeneity of these protests and evolution of key issues associated with COVID-19 add new challenges to the problem of predicting these protests. In addition, since the full trend of protest counts provides more information, we wanted to predict not only a binary protest outcome, but also the overall trend of protests over time. To explore these relationships, we used data from eight countries in Western Europe and four regions from the United States (i.e., West, South, Midwest, and Northeast). We first explore the relationships between our data streams and COVID-19-related protests. We build forecasts using logistic regression and random forest methods for a binary protest outcome and compare our results to models from the literature. For forecasting protest counts, we used Poisson regression and random forest. All model performance was evaluated against a naïve forecast. In the following sections, we will first describe the data streams and statistical methods we used to examine the data and create forecast models. We then present both a narrative of our exploratory analysis of the data streams and the forecast results in the Results section. Finally, we provide a discussion of our results and their limitations.

\section*{Materials and methods}
\subsection*{Scope}
\begin{figure}[!h]
\includegraphics[width=.5\textwidth, center]{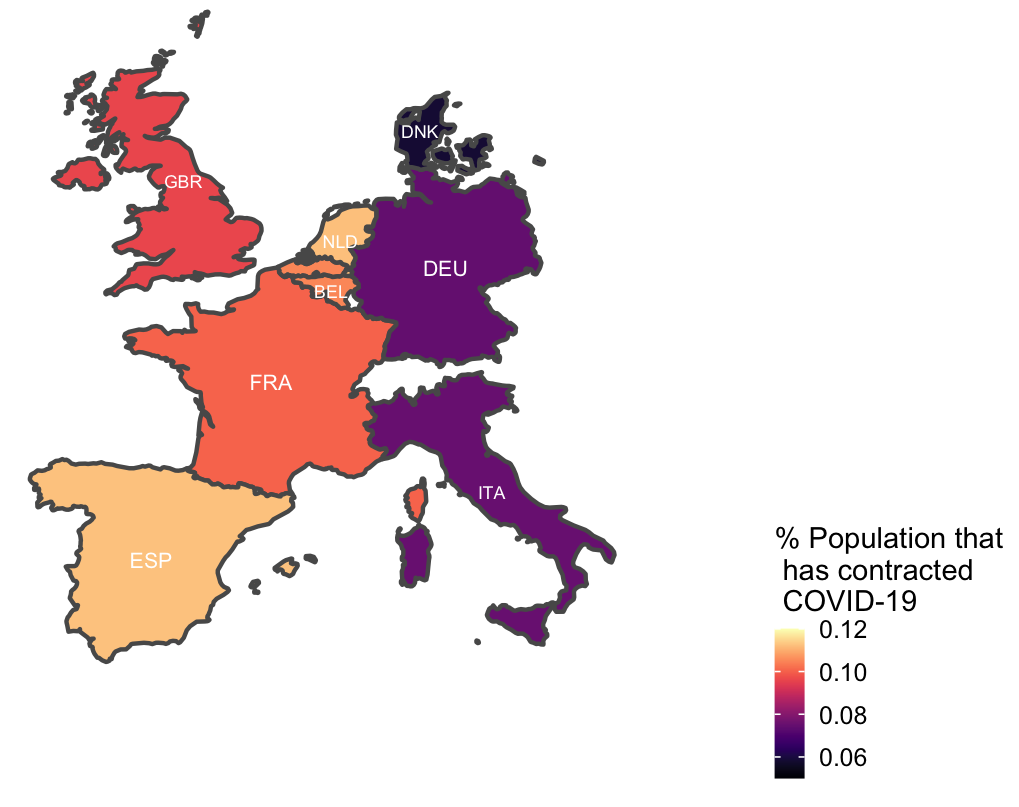}
\includegraphics[width=.75\textwidth]{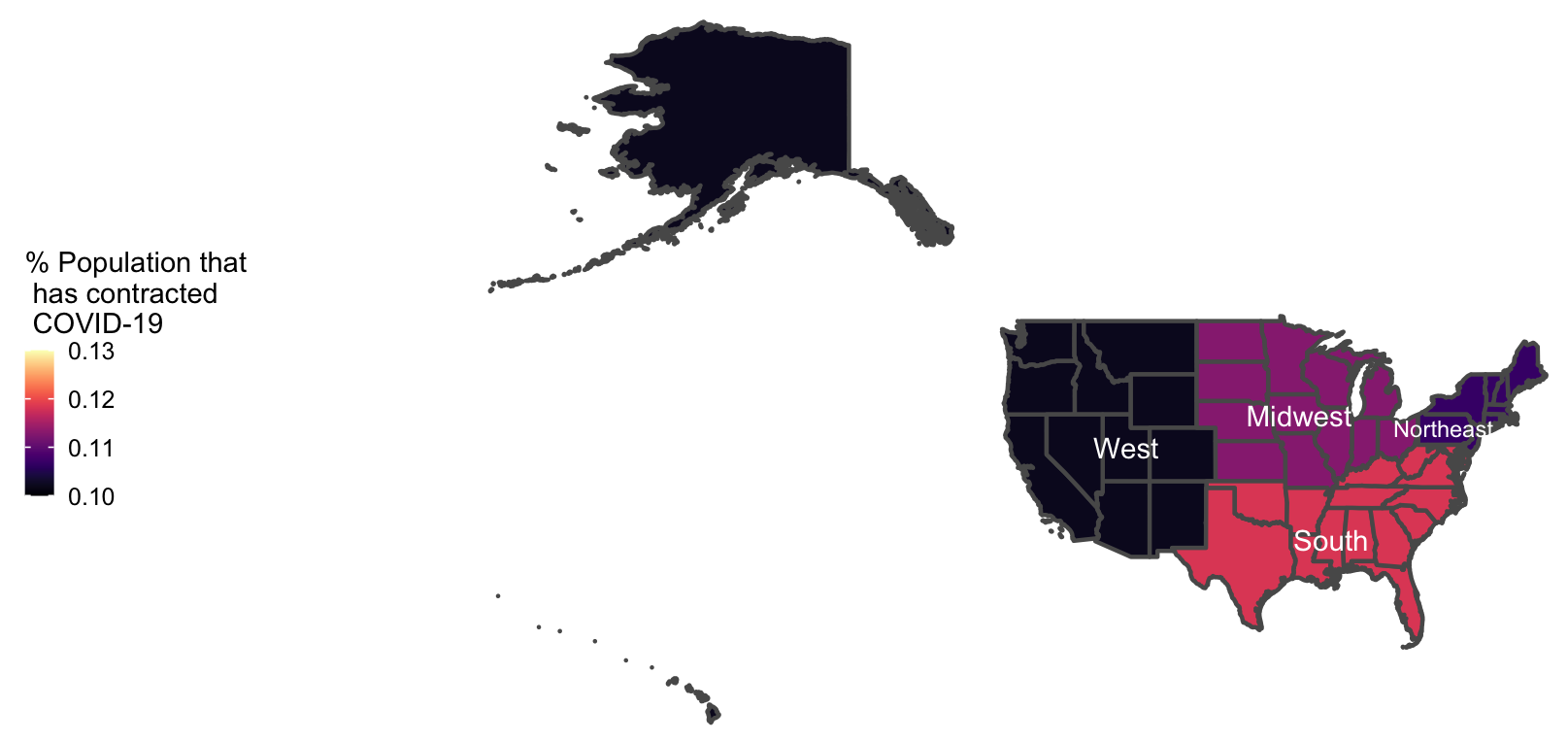}
\caption{{\bf August 2021: Proportion of Population that has had COVID-19}
Each map shows the proportion of the given region or country that has contracted COVID-19, assuming no reinfection (total positive tests / population) \cite{jh2021}. Note that the United States and Europe maps have a different scale to the color legend. In the United States, the South has had the highest proportion of cases, while in Western Europe, Spain and the Netherlands have had highest proportion of cases.}
\label{fig1}
\end{figure}

In order to compare the spatial differences in COVID-19-related protests and their drivers, we first chose regions that are peer countries as well as geographically adjacent to each other. Specifically, we examined eight countries in Western Europe and the four regions of the United States from January 2020 to July 2021. Abbreviations for the regions are included in Table \ref{table1}. Note that countries in Western Europe enacted stronger public health policies, such as more extended and stricter lockdown measures, than in the United States. Within Western Europe, Denmark has had the lowest COVID-19 incidence, while the Netherlands and Spain have had the highest (Fig \ref{fig1}). Generally, countries that are further west have higher COVID-19 incidence, and the regions with the lowest incidence are in Europe.

Given the extensive heterogeneity in regards to policies and healthcare impacts within the United States, we decided to stratify the U.S. into four different regions (Northeast, Southwest, Midwest, and South) based on similarities in trends. Specifically, the Northeast had the most extreme first wave, while the South and Midwest had the most extreme second and third waves, respectively. Generally, the South and Midwest have had the highest rates of COVID-19 infection throughout the pandemic.

\begin{table}[!ht]
\centering
\caption{
{\bf Region name abbreviations}}
\begin{tabular}{|l|l|}
\hline
{\bf Name} & {\bf Abbreviation} \\ 
\thickhline
Belgium & BEL \\ \hline
Germany & DEU \\ \hline
Denmark & DNK \\ \hline
Spain & ESP \\ \hline
France & FRA \\ \hline
Great Britain & GBR \\ \hline
Italy & ITA \\ \hline
Netherlands & NLD \\ \hline
U.S. Midwest & MW \\ \hline
U.S. Northeast & NW \\ \hline
U.S. West & W \\ \hline
U.S. South & S \\ \hline
\end{tabular}
\label{table1}
\end{table}

\subsection*{Data}
\subsubsection*{The Armed Conflict Location and Event Data Project}
Data about political violence and protest events were obtained from the Armed Conflict Location and Event Data Project (ACLED) from January 1st 2020 to July 30th 2021 for Europe and January 1st, 2020 to June 25th, 2021 for the United States ~\cite{acled2021}. These data include the location, date, type, violence level, actors, and description the each event. We included all events in our regions of interest, and selected events related to COVID-19 by filtering events that included the words "coronavirus" or "covid" in their description. We chose to run our analysis on the weekly scale for temporal consistency with other data sets. To aggregate protest data to this scale, we summed COVID-19 protest counts by week. For the United States, state level data was aggregated up to the regional level by taking the sum of all state level protests in a given region and week. ACLED uses a variety of sources (e.g., traditional media, government reports, local organizations, social media) to generate their data. While some of their sources may be biased in specific regions, they perform quality control of both the sources and their data \cite{ACLEDGuides2021}. 

\subsubsection*{Johns Hopkins COVID-19 Data}
For COVID-19 health outcomes data, we used the cumulative COVID-19 case and death data from the COVID-19 Data Repository by the Center for Systems Science and Engineering at Johns Hopkins University that was published within the OxCGRT data from January 1st 2020 to August 16th, 2021 ~\cite{jh2021}. We converted these metrics to daily cases and death counts. For temporal consistency across data streams, we then took the weekly sum of daily COVID-19 case and death counts. For the United States, state level data was aggregated up to the regional level by taking the sum of weekly cases and deaths.

\subsubsection*{Oxford COVID-19 Government Response Tracker Data}
The Oxford COVID-19 Government Response Tracker (OxCGRT) has been tracking government policy changes since January 1st 2020 in over 180 countries~\cite{oxcgrt2021}. They present data as daily indicator of government action for various policies. We examined their data on eight closure policy indicators, four economic policy indicators, and eight health policy indicators (Table~\ref{table2}, full codebook at \cite{oxcgrt_git2020}). We did not use the data on three vaccine policy indicators and one miscellaneous indicator due to our time frame of interest and the lack of consistent meaning of the miscellaneous indicator. The majority of the policy indicators are ordinal metrics, where higher numeric values indicate higher levels of government action. The data also included an additional metric or flag for each policy indicator that reported the geographic scope of the policy (targeted (i.e., sub national) or general (i.e., national)). However, the flags were nonspecific about the targeted region of the policy, and therefore, we did not adjust policy indicator values based on geographic scope as there was no way to determine the population size that experienced the targeted government action. There was limited missing data, however, any missing data in the policy indicators was conservatively given a value of $0$, indicating that there was no policy at that time. In addition, we explored the four calculated indices that are aggregations of the individual policy indicators: overall government response index (all indicators, see Table \ref{table2}), containment and health index (all closure and health indicators), stringency index (all closure and one health indicator), and economic support index (all economic indicators). All indices are on a scale of $0-100$. For temporal consistency across data streams, we took the weekly average of policy indicators and indices. For the United States, state level data was aggregated up to the regional level by taking a weighted mean by population of the policy indicators and indices. 

All data is related to the day the given policy was implemented, not announced \cite{oxcgrt_git2020}. In addition, the strength level of a policy in a given geographic region is the highest strength that exists, even if that policy strength only exists in a sub-region. Data also is checked by reviewers, and therefore, some gets retroactively changed, although we used the most recently available data as of August 16th 2021 to limit this possibility. 

\begin{table}[!ht]
\centering
\caption{
{\bf Specific policy indicators explored from the Oxford COVID-19 Government Response Tracker Data}}
\begin{tabular}{|l|l|l|}
\hline
{\bf Closure} & {\bf Economic} & {\bf Health}\\ 
\thickhline
School closing & Income support & Public information campaigns \\ \hline
Workplace closing & Debt/contract relief & Testing policy \\ \hline
Cancel public events & Fiscal measures$^*$ & Contact tracing \\ \hline
Restrictions on gatherings & International support$^*$& Emergency investment in healthcare$^*$ \\ \hline
Close public transport &  & Investment in vaccines$^*$ \\ \hline
Stay at home requirements &  & Facial coverings \\ \hline
Restrictions on internal movement &  & Vaccination policy \\ \hline
International travel controls &  & Protection of elderly people \\ \hline
\end{tabular}
\begin{flushleft} $^*$ Indicator measured in U.S. dollars (USD) rather than an ordinal metric
\end{flushleft}
\label{table2}
\end{table}

\subsubsection*{Google Trends Data}
Google Trends provides data on the relative search term volume for any search query of interest~\cite{gtrends}. For each term with sufficient data, the search volume is scaled from 0-100 in the time period of interest. We examined search terms related to misinformation in the following conceptual groupings: general (e.g., 'fake news'), COVID-19 (e.g., 'covid hoax'), lockdowns (e.g., 'anti-lockdown'), school opening/closure (e.g., 'open schools'), masks (e.g., 'anti-mask'), vaccines (e.g., 'anti-vax'), and economic terms (e.g., 'open stores') (full list of terms in \nameref{S1_Table2} Table 2). Some of these groupings contain words not specifically associated with misinformation (e.g., 'pfizer' in the vaccine grouping). However, since for any given topic, misinformation tends to be more popular than evidence-based information, we assumed that search increases in the term in general may correspond to increases in misinformation as well~\cite{Wang2019}. For countries where English is not the primary language (i.e., Italy, France, Spain, Germany, Denmark, Belgium, and Netherlands), we used the googleLanguageR R package, version 0.3.0, to access the Google Translate API on the Google Cloud Platform \cite{gtranslate}. After the translations, we replaced words that were mistranslated, as many countries have created new words throughout the COVID-19 pandemic~\cite{oxfordlanguage}. Using the gtrendsR R package, version 1.4.8.9000, we pulled Google Trends data from January 5th, 2020 to August 1st, 2021 for all search terms. Some search terms did not have data in certain countries, possibly due to mistranslation or limited searches. We then averaged the search volumes over time for the terms in each conceptual grouping. This resulted in seven metrics of misinformation, each associated with the original conceptual grouping. For the United States, state level data of these seven metrics was aggregated up to the regional level by computing a weighted mean by population.

\subsection*{Statistical Methods}
\subsubsection*{Data Exploration}
We analyzed all data at the weekly level, where public health orders, COVID-19 outcomes, and protests data were aggregated up to that temporal scale. We include data from late January 2020 to July 2021 for Europe and from February 2020 to June 2021 for the United States, though some regions have a later start date based on the timing of the first COVID-19-related protest. We first examined the relationships across these data streams. To quantify the relationships, we ran Granger Causality Tests for $1-4$ week time lags between COVID-19-related protests and all variables in the data streams, except public health variables that were measured in U.S. dollars (USD). The Granger Causality Test assess whether a lagged explanatory variable adds information to a univariate autoregression of the outcome through both the explanatory variable's t-statistic and the F-test between the regressions. This test, rather than suggest causality, evaluates whether the data stream at the given time lag has significant capacity to forecast the protests. The tests were performed using the lmtest R package, version 0.9.38.

\subsubsection*{Forecasting Methods}
A forecasting model was fit for each individual geographic region. We performed two sets of forecasts for each region: one that predicted a weekly binary protest outcome and one that predicted the weekly protest count. We created a binary protest outcome as follows, 
\[ Y^b_t = \begin{cases} 
      1 & Y_t > Q(Y, 0.75) \\
      0 & Y_t \leq Q(Y, 0.75)
   \end{cases}
\]
where $Y_t$ is the weekly number of protests at week $t$ and $Q(Y, 0.75)$ is the third quartile of the protest counts throughout the complete time frame of our data. We chose the third quartile as the threshold because we were concerned with predicting times of high protest counts correctly. For the binary outcome forecasts we used logistic regression and random forest, while for the count outcome we used Poisson regression and random forest machine learning methods. For both outcomes, we performed $i$ week forecasts, $i \in \{1,2,3\}$. Logistic regression models the logarithm of the odds of a binary response, while the Poisson regression models the logarithm of the expected value of a Poisson random variable, which is a distribution used to express count data. The random forest model is an ensemble process that builds an explicit number of decision trees with randomly selected predictors and then averages the results to obtain the final forecast values. 

For all models, we started with an initial training period of the earliest protest data for a given region (spanned from late January to early March 2020), to the end of October 2020. For an $i$ week forecast, we retrained the model every $i$ weeks. After the initial $i_0$ week forecast, for every next $i_j$ week forecast, the train period starts and ends $i$ weeks later than the $i_{j-1}$ forecast train period. This is a rolling training period such that the training period is the same size each time the model is retrained. We used a rolling training window because exploratory analysis showed that the relationships between the predictors and protests changed over time due to the fluctuating state of information and public health orders throughout the pandemic (e.g., vaccine related predictors are non existent in 2020, but are in 2021). The total rolling testing period spans from November 2020 to July 2021 for Europe and November 2020 to June 2021 for the United States. Throughout this section, $Y_t$ indicates the protest count at week $t$, $F_t$ indicates the forecast protest count at week $t$ and $Y^b_t$ and $F^b_t$ have the equivalent meanings for the binary outcome.

For each $i_j$ forecast, all $29$ predictors had an $i$ week time lag (full list of predictors in \nameref{S1_Table3} Table 3). Feature selection was performed on each training set. We used logistic and Poisson regression with stepwise Akaike Information Criterion (AIC), performed with the MASS R package version 7.3.54, to select the best model predictors. However, with a large number of predictors, especially for data with fewer observations, stepwise AIC tends to overfit the models ~\cite{James2013}. Thus, we first selected the eight predictors (i.e., the maximum number of predictors to include in a model based on our data) with the smallest p-values resulting from Granger Causality Test at an $i$ week lag. The logistic and Poisson regressions were subsequently trained on the set with the selected predictors.

For random forest, we first selected all predictors with a p-value $< 0.05$ from the Granger Causality test at an $i$ week lag. If there were less than $9$ predictors (we wanted at least three predictors in each tree, using $\sqrt{\text{total input predictors}}$) that had a p-value $< 0.05$, we selected the $9$ predictors with the lowest p-values. For the random forest model, we used a number of trees such that the model error stabilized, namely $500$ trees ~\cite{James2013}. Each tree used $\sqrt{\text{total input predictors}}$ predictors. Regressions were performed with stats R package, version 4.1.0, and random forest models were performed using the randomForest R package, version 4.6.14. 

\subsubsection*{Forecast Evaluation}
We evaluated the true positive rate (TPR) (fraction of $Y^b_t =1$ correctly predicted as $F^b_t = 1$), the true negative rate (TNR) (fraction of $Y^b_t =0$ correctly predicted as $F^b_t = 0$) and the balanced accuracy, BAC = $\frac12$(TPR + TNR), to assess our binary outcome forecasts ~\cite{Kallus2014, Bahrami2018}. Note that the TPR is also known as sensitivity. The BAC metric helps account for the uneven cutoff of our binary outcome. Note that due to our cutoff designation, $Y^b_t = 0$ for $75\%$ of our full data period. Therefore, if we forecast $F^b_t = 0$ for every $t$, the overall error rate will most likely be greater than $0.5$, however, the BAC $= 0.50.$ In addition, a forecast without relationship to the outcome will achieve a BAC $= 0.50$ on average by statistical independence. We also calculated these metrics for the naïve $i$ step forecast (i.e., the protest status (high/low) $i$ weeks away is the same protest status as today) to compare our models to.

To evaluate our forecasts of protest counts, we used Pearson's $R^2$ and a version of Mean Absolute Scaled Error (MASE) as our primary model evaluation metrics~\cite{Hyndman2006}. Pearson's correlation squared, $R^2$, evaluates how well the forecast matches the pattern of the true values. The value ranges from $0$ to $1$, where $1$ represents perfect positive or negative correlation, and $0$ represents a complete lack of relationship. We calculated $R^2$ between the forecasts and data, as well as between forecast shifted up $1-2$ weeks and the data to determine whether the model generally captured the protest trend, but not at the exact right time. The traditional MASE is calculated as follows~\cite{Hyndman2006}:
\begin{align*}
    e_k &= |Y_k - F_k| \\
    N_{_\text{MAE}} &= \frac{1}{T-1}\sum_{t=2}^T |Y_t - Y_{t-1}| \\
    \text{MASE} &= \text{mean}(\frac{e_1}{N_{_\text{MAE}}}, \ldots, \frac{e_K}{N_{_\text{MAE}}}) = \frac{\frac{1}{K}\sum_{k=1}^K e_k}{N_{_\text{MAE}}}
\end{align*}
where $e_k$ is the forecast error from forecast period $k$ and $N_{_\text{MAE}}$ is the mean absolute error (MAE) of the one step naïve forecast method over the time period of interest. We have changed this slightly by calculating the forecast error at each time point and altering the naïve model comparison to an $i$ step naïve method for an $i$ week forecast. Therefore, we have the following metric:
\begin{align*}
    e_t &= |Y_t - F_t| \\
    N_{i_{\text{MAE}}} &= \frac{1}{T-i}\sum_{t=1+i}^T |Y_t - Y_{t-i}| \\
    \text{MASE}_i &= \text{mean}(\frac{e_1}{N_{i_{\text{MAE}}}}, \ldots, \frac{e_T}{N_{i_{\text{MAE}}}}) = \frac{\frac{1}{T}\sum_{t=1}^T e_t}{N_{i_{\text{MAE}}}}
\end{align*}
for an $i$ week forecast. An $\text{MASE}_i = 1$ indicates that the MAE of the $i$ week forecast is equal to the MAE of the naïve $i$ step forecast method, while an $\text{MASE}_i < 1$ indicates the $i$ week forecast has a smaller error, and an $\text{MASE}_i > 1$ indicates that the $i$ week forecast has a larger error. All analyses were performed in R version 4.1.0.

\section*{Results}

\subsection*{Relationships between COVID-19 Health Outcomes, Public Health, Misinformation, and Protests}
 
\begin{figure}[!h]
\includegraphics[width=.85\textwidth, center]{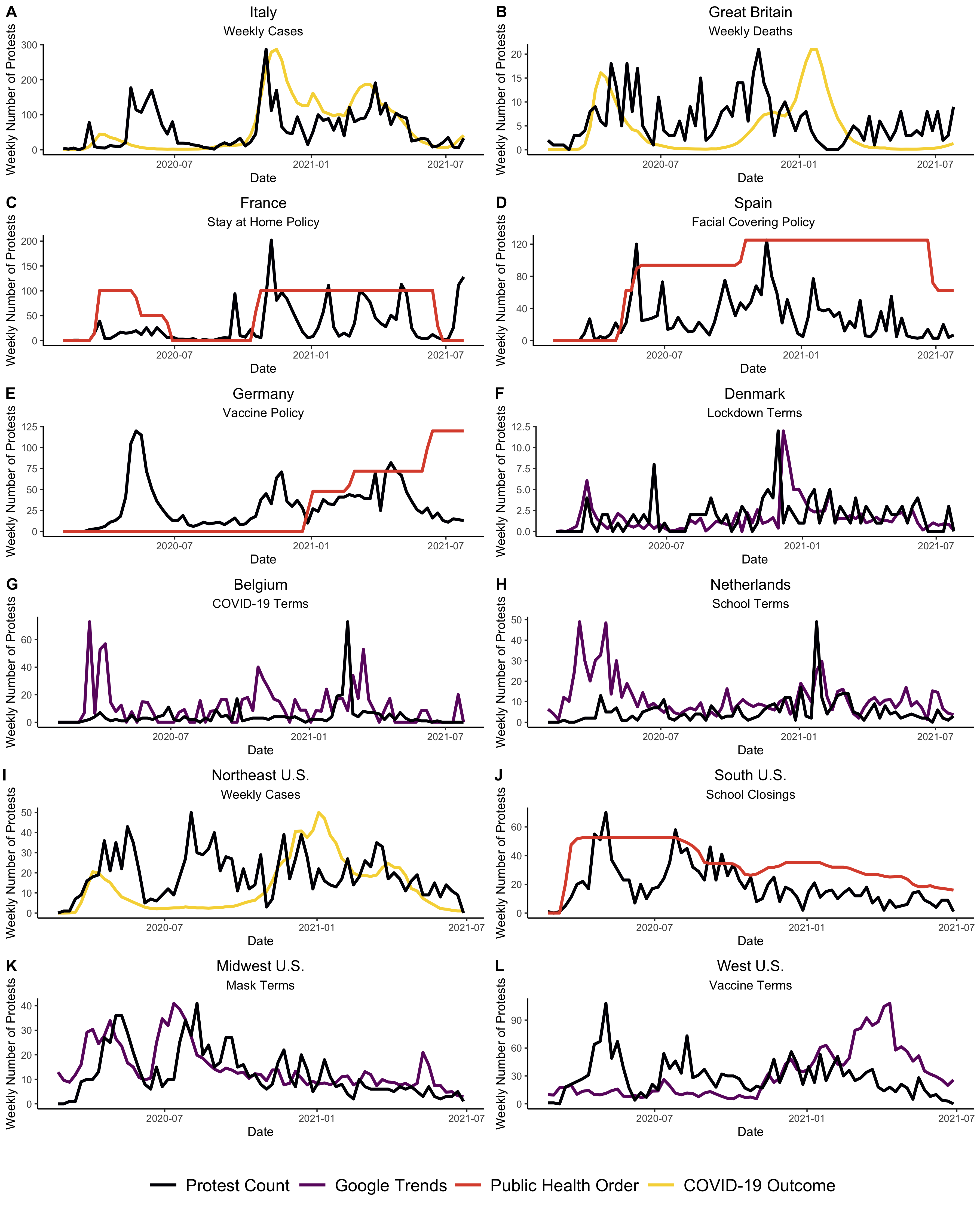}
\caption{{\bf Relationship between weekly COVID-19 cases and deaths, public health orders, and Google Trends.}
Panels A-L show the weekly COVID-19-related protests counts for all regions of interest along with one variable from the data streams (Europe at the top, United States at the bottom). The vertical axis depicts the number of weekly protests for all panels, the data streams were scaled to match the magnitude of the protest data. Note that the vertical axis scale is different for all the figures shown. The color of the additional trend line represents the type of data, while the subtitle defines the specific variable from that data stream being visualized. The relationships between the data streams and COVID-19 protests shown are indicative of the trends across all the regions.}
\label{fig2}
\end{figure}

Protests follow similar trends in Europe with the majority of countries having two waves of protests in 2020, the first wave peaking in May and the second wave in November (Fig \ref{fig2} A-H). The two countries that do not follow this trend are the Netherlands and Belgium, where there was only one protest peak that occurred in early 2021. Based on the protest description in the ACLED data, these protests peaks were associated with renewed lockdowns and a curfew in the Netherlands and restrictions on food and hospitality businesses in Belgium. The protest trends in each country become less similar in 2021: some countries have very few protests (i.e., Belgium, Netherlands), others have one protest wave (i.e., Italy, Germany), and others are very variable throughout 2021 (i.e., France, Spain).

Regional groupings of the United States also showed some similarities between regions (Fig \ref{fig2}I-L). All four regions (i.e., Midwest, Northeast, South, and West) have two clear peaks in 2020. In the South and the West regions, the second peak is smaller, whereas in the Northeast and the Midwest regions, the second peak is slightly larger. Notably, protests in all regions, except in the Northeast, significantly taper off after the two peaks and steadily decrease into 2021. In the Northeast, there are clear fluctuations in the number of protests continuing into 2021 (Fig \ref{fig2}I). 

In the European countries with a two wave pattern, there was a relationship between protests and COVID-19 weekly cases and deaths. The first peak in cases occurs before the first peak in protests and the second peak in cases occurs concurrently with the second peak of protests. However, the first peak in deaths occurs concurrently with the first wave of protests and the second peak of deaths occurs after the second peak of protests (Figs \ref{fig2}A \& \ref{fig2}B). This demonstrates that the relationship between COVID-19 outcomes and protests changes over time, with there being a much more substantial time lag at the beginning of 2021. This trend continues with the relationship between public health orders and protests; the first wave of protests has a more substantial time lag with public health order strength than the second wave (Fig \ref{fig2}C). Some public health orders increase in strength right before an increase in protests, such as facemask or vaccine policy. This does not indicate causation, however, there may be some non-linear relationship between these orders and protests (Figs \ref{fig2}D \& \ref{fig2}E). These time lags fit with the general of trajectory of COVID-19 policy at the beginning of the pandemic: most countries enacted very strict lockdowns, which prohibited individuals from activities, like protests, until the restrictions were loosened. 

The averaged Google Trends misinformation metrics were also related to COVID-19-related protests in Europe, however, they did not have the same time lag relationship as the other two data sources (Figs \ref{fig2}F-H). Different metrics were most correlated with protests in each country, however, COVID-19, lockdown, school, and mask terms tended to have the highest correlations. While the Google Trends data is correlated, the search volumes have a high magnitude at the beginning of 2020 that does not match the magnitude of protests, as attention towards COVID-19 was incredibly high at the beginning of the pandemic (Figs \ref{fig2}G \& \ref{fig2}H). 

Public health orders loosely followed a similar trajectory as weekly protest counts in some regions of the United States. For example, in the South, school closing public health orders experienced fluctuations along with the number of protests (Fig \ref{fig2}J). There is also a similar time lag present in the relationships between public health orders and protests in 2020 in the United States as seen in Europe. In the Midwest, protest counts closely aligned with Google Trends data on mask-related terms (Fig \ref{fig2}K). In this region, the Google Trends counts for mask terms peaked before the protests peak and subsequently declined along with the protests, indicating some amount of time lag between Google searches and protests in the United States. Generally, the South and the Northeast had high correlations between economic search terms and protests. Regions also tended to have high correlations between vaccine terms and and protests in early 2021 (Fig \ref{fig2}L). Like the trends seen in Europe, the relationship between protests and other variables are not consistent over time. For example, in the Northeast, correlation between protests and COVID-19 cases varies significantly throughout the time frame (Fig \ref{fig2}I). 

\begin{figure}[!h]
\includegraphics[width=\textwidth]{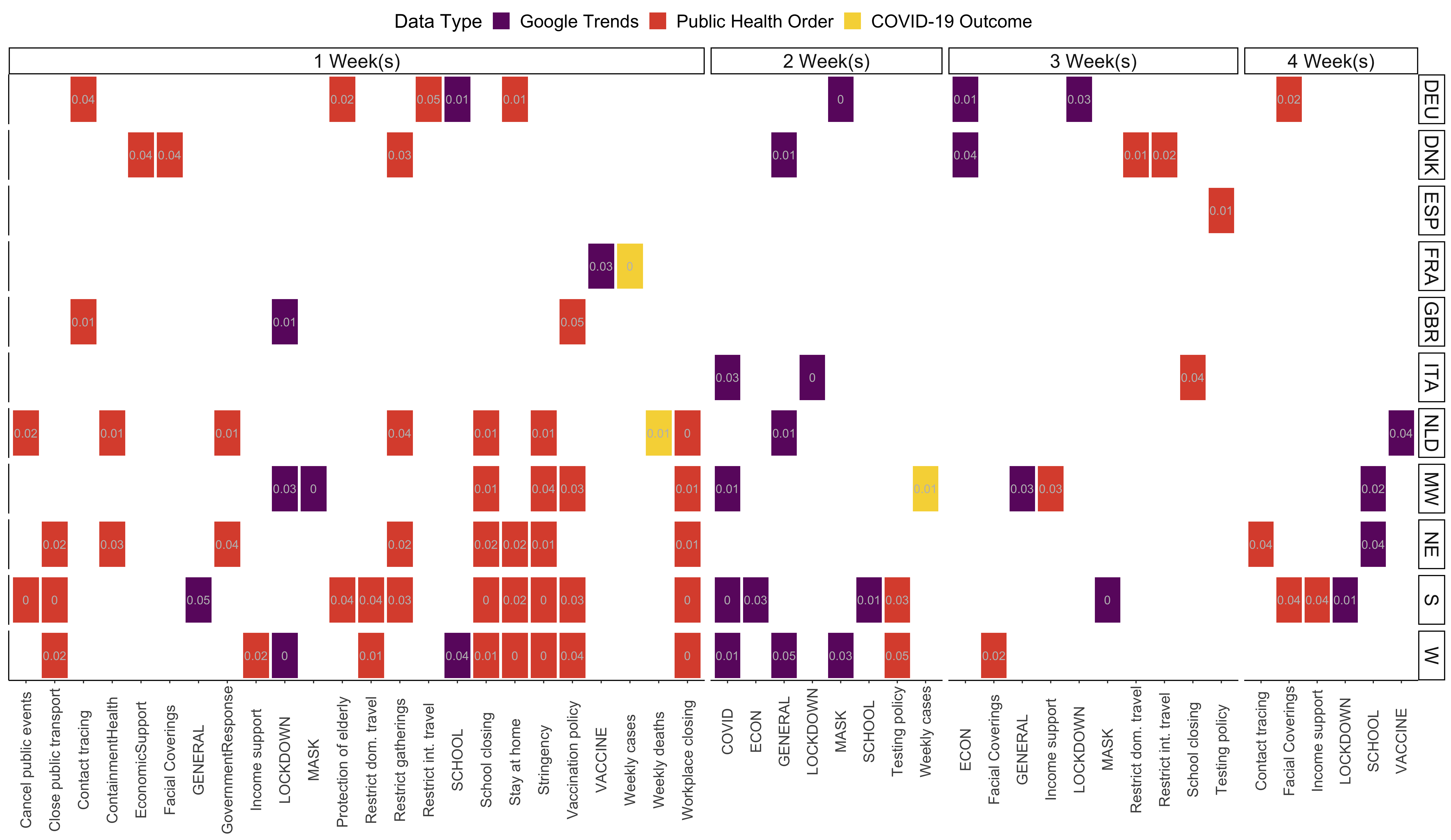}
\caption{{\bf Granger Test associations}
This plot depicts the variables that were significant for each region at the $\alpha = .05$ level for the Granger Causality Test, at the time lag with the lowest p-value for each variable. The horizontal axis lists all significant variables across regions. Each colored box demonstrates that a specific variable was significant at the $\alpha = .05$ level in a given region (row) and time lag (column). The color of the box depicts the specific data stream the variable comes from, and the p-value rounded to the second decimal place is printed in the box.}
\label{fig3}
\end{figure}

Multiple variables had significant forecasting COVID-19 protest potential at the $\alpha = .05$ level for the European countries (Fig \ref{fig3}). Public health orders tended to be significant at a one-week time lag, specifically closure and health policy. The significant time lag for Google Trends was variable across European countries, although the most were significant at a two-week time lag. Overall, most variables were significant at the one-week time lag, and very few were significant at a four-week time lag. In addition, some countries had many more significant variables than others. Germany, Denmark, and the Netherlands all had over five significant variables, while Belgium (not pictured in Fig \ref{fig3}) had none. Therefore, despite seeing common trends in visualizations, each of the European countries analyzed have distinct relationships between the various data streams and COVID-19-related protests. Across all countries, public health orders and Google Trends aggregated metrics tend to have a forecasting potential, while COVID-19 outcomes were only significant in France and the Netherlands.

In the United States, many variables had significant forecasting potential at the $\alpha = .05$ level, particularly when using a one-week time lag (Fig \ref{fig3}). Similar to Europe, public health orders were generally significant at a one-week lag. Public health orders regarding school closures and workplace closures, as well as the stringency index were significant for all regions at a one-week time lag. Stay-at-home and public transport related health orders were significant at a one-week lag for all regions except the Midwest, and vaccination-related public health orders were significant at the one-week lag for all regions except the Northeast. Notably, the Google Trends terms were largely significant at a two-week time lag. All four regions had a substantial number of significant variables, largely from the public health order and Google Trends data streams. However, the regions do exhibit some differences in which variables are significant, indicating unique relationships by region. However, regions in the United States tended to have many more predictors significantly related to protests than the majority of the European countries. 

 Thus, in Europe and the United States, COVID-19 policy and misinformation does appear to contribute, and has forecasting potential, with COVID-19-related protests, however, the fact that the majority of variables had the lowest p-value at one-week time lag suggests that many of these data streams may not have long-term forecast ability.  This fact makes intuitive sense considering the rapid evolution of COVID-19, that is, public health orders and misinformation circulating today should not be related to protest activity over a month later. 

\subsection*{Forecasting}
\subsubsection*{Binary Protest Outcome}

\begin{figure}[!h]
\includegraphics[width=.85\textwidth, center]{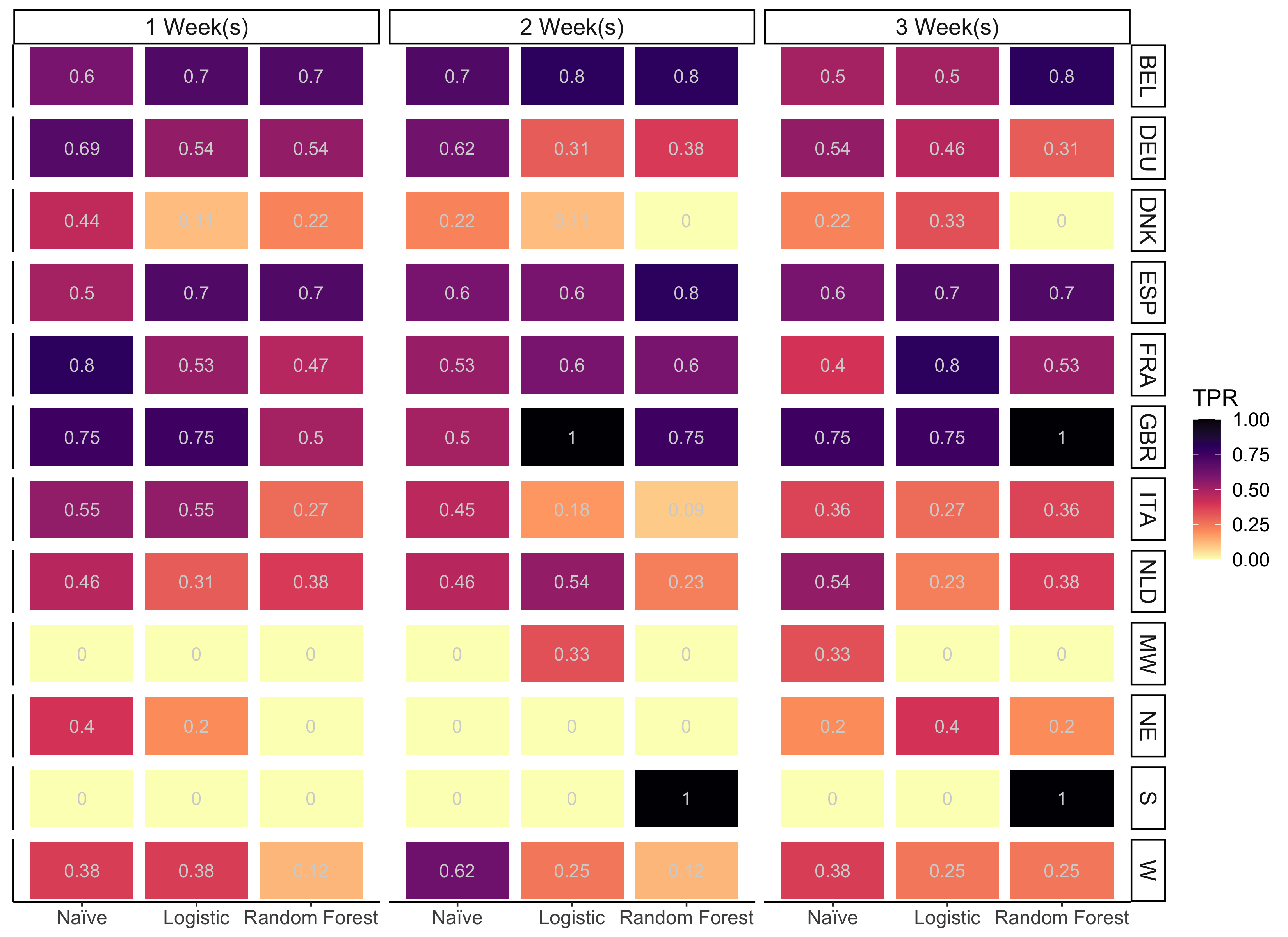}
\caption{{\bf True positive rate from binary response forecasts.}
This plot shows the true positive rate (TPR) from forecasts of a binary protest response. The horizontal axis labels the model method, where both logistic regression and random forest can be compared to the baseline naïve model. Each colored box depicts the TPR for a forecast method at a specific forecast time frame (column) and region (row). The boxes are colored and labeled with the TPR (darker color: better TPR).}
\label{fig4}
\end{figure}

\begin{figure}[!h]
\includegraphics[width=.85\textwidth, center]{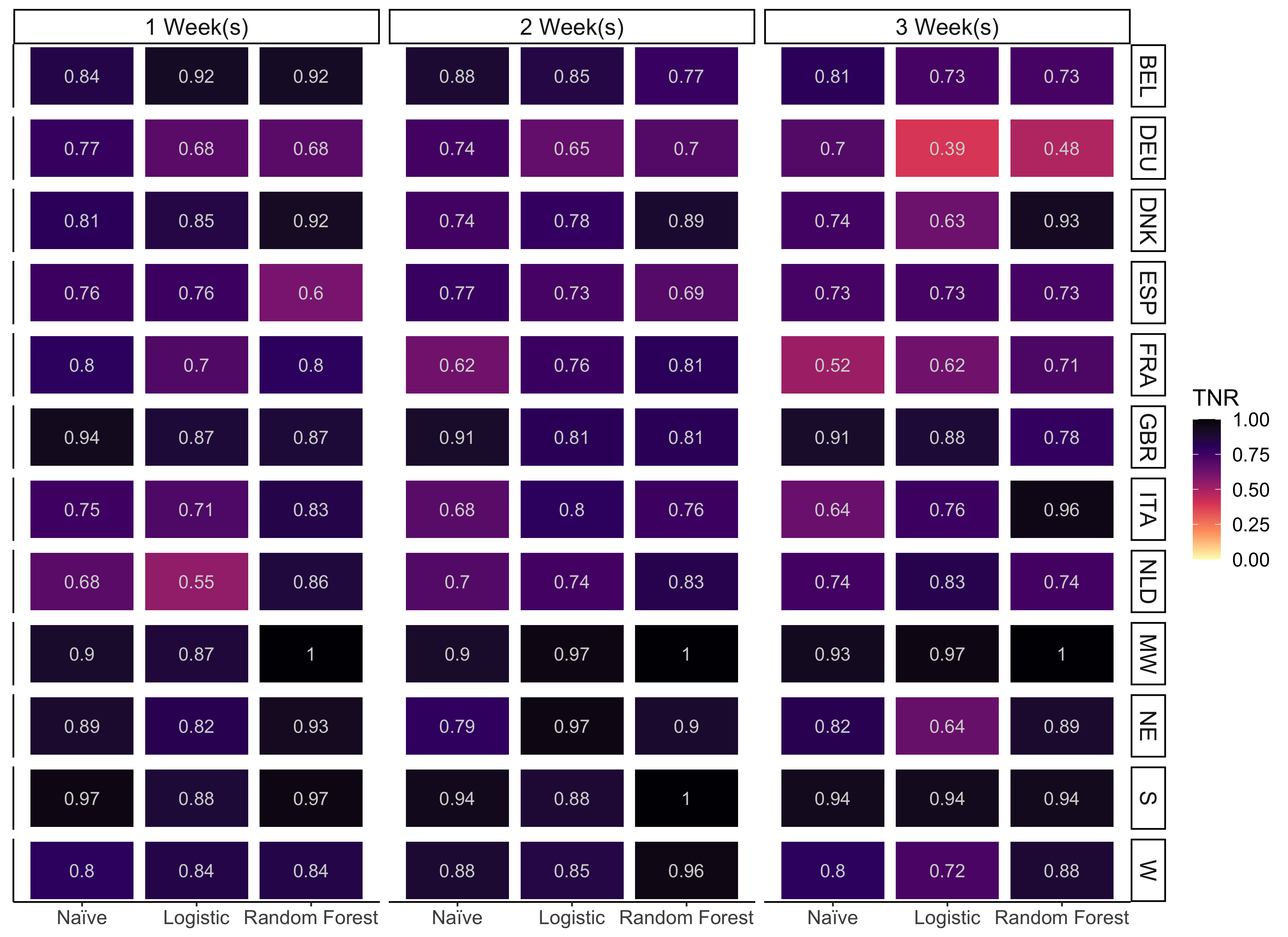}
\caption{{\bf True negative rate from binary response forecasts.}
This plot shows the true negative rate (TNR) from forecasts of a binary protest response. The horizontal axis labels the model method, where both logistic regression and random forest can be compared to the baseline naïve model. Each colored box depicts the TNR for a forecast method at a specific forecast time frame (column) and region (row). The boxes are colored and labeled with the TNR (darker color: better TNR).}
\label{fig5}
\end{figure}

The logistic regression and random forest had varying success by country in Europe. The forecasts outperformed the naïve forecast TPR in Belgium, Great Britain, and Spain at all forecast periods (Fig \ref{fig4}). The forecasts also had higher TPR than the naïve forecast for France at two and three week forecasts. Germany, Denmark, Italy, and the Netherlands had very low forecast TPRs, which got lower with longer forecast periods. The forecasts for most countries had relatively high true negative rates and the majority of these forecast outperformed the naïve forecast (Fig \ref{fig5}). Great Britain, Spain, and Germany are the only countries where the forecasts consistently had lower TNRs than the naïve forecast.


\begin{figure}[!h]
\includegraphics[width=.85\textwidth, center]{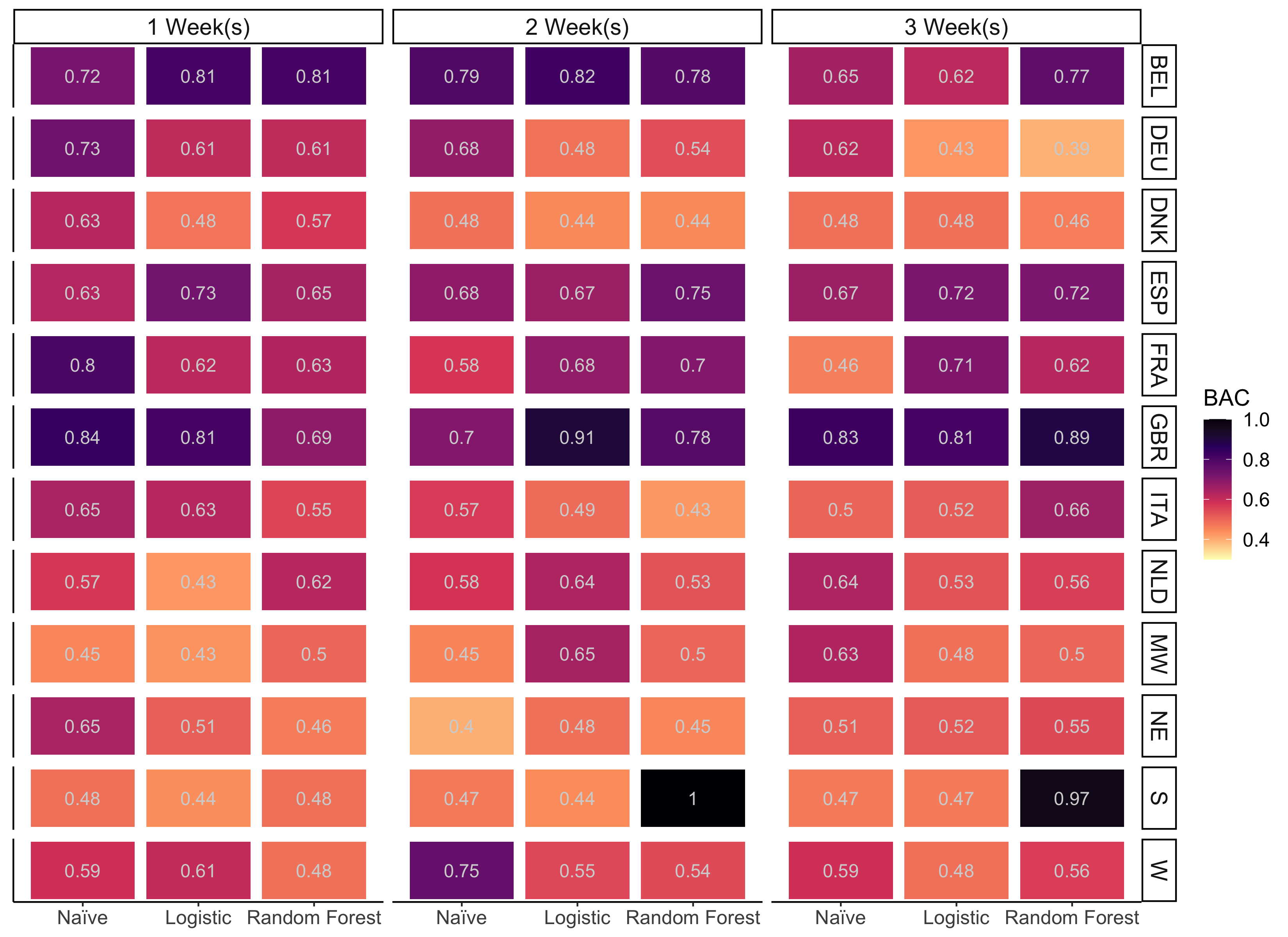}
\caption{{\bf Balanced accuracy from binary response forecasts.}
This plot shows the balanced accuracy (BAC) from forecasts of a binary protest response. The horizontal axis labels the model method, where both logistic regression and random forest can be compared to the baseline naïve model. Each colored box depicts the BAC for a forecast method at a specific forecast time frame (column) and region (row). The boxes are colored and labeled with the BAC (darker color: better BAC).}
\label{fig6}
\end{figure}

Since all forecasts across countries had relatively high TNRs, the countries with forecasts with high TPRs also have high BACs (Fig \ref{fig6}). Therefore, the BACs for at least one forecasting model for Belgium, Great Britain, and Spain outperformed the naïve model across all time frames, except at a one-week forecast in Great Britain. Great Britain had especially high BACs at the two- and three-week forecasts. At least one forecasting model BAC outperformed the naïve forecast for the two- and three-week forecasts for France, the one- and two-week forecast for the Netherlands, and the three-week forecast for Italy. The forecasts for Germany and Denmark all performed poorly as measured by BAC, having values around or lower than $0.5$, which is the resulting BAC of a forecast with no relationship with the outcome. 

Forecasting using a binary protest response was less successful in the U.S. regions than in Europe, excluding the South. One reason for this might be that the U.S. regions generally had more "low protest" counts than "high protest" counts in the test data, which can help explain the low true positive rate for most of the forecasting. In the South, the random forest TPR performed exceptionally well at the two- and three-week lag, significantly outperforming the others (Fig \ref{fig4}). However, there was only one "high protest" week in the test set in the South, meaning the forecast only had to predict that week correctly to achieve TPR $=1$. In the Midwest, the logistic regression TPR outperformed the naïve TPR at the two-week forecast, however the naïve forecast TPR was higher than both three-week model forecasts. In the West, the naïve forecast TPR outperformed the models at all forecast periods. In the Northeast, only the three-week logistic regression forecast TPR outperformed the naïve forecast TPR. The forecasts for all regions tended to have relatively high TNRs consistent with the low forecast TPRs (Fig \ref{fig5}). The random forest forecast TNRs consistently outperformed the naïve forecast at all forecast times and regions.

The BACs for the U.S. regions (excluding the South) were similarly low to the TPRs (Fig \ref{fig6}). The random forest outperformed the other methods at the two- and three-week forecasts in the South and at the three-week forecasts in the Northeast. The naïve approach outperformed the three-week forecasts in the Midwest and the West, the two-week forecast in the West, and the one-week forecast in the Northeast. The logistic regression outperformed the others at the one-week forecast in the West and the two-week forecast in the Midwest and Northeast. However, many of the forecasts for these regions had BAC values of $0.5$, or less, indicating a forecast with no relationship to the outcome.

\subsubsection*{Protest Count Outcome}

\begin{figure}[!h]
\includegraphics[width=.85\textwidth, center]{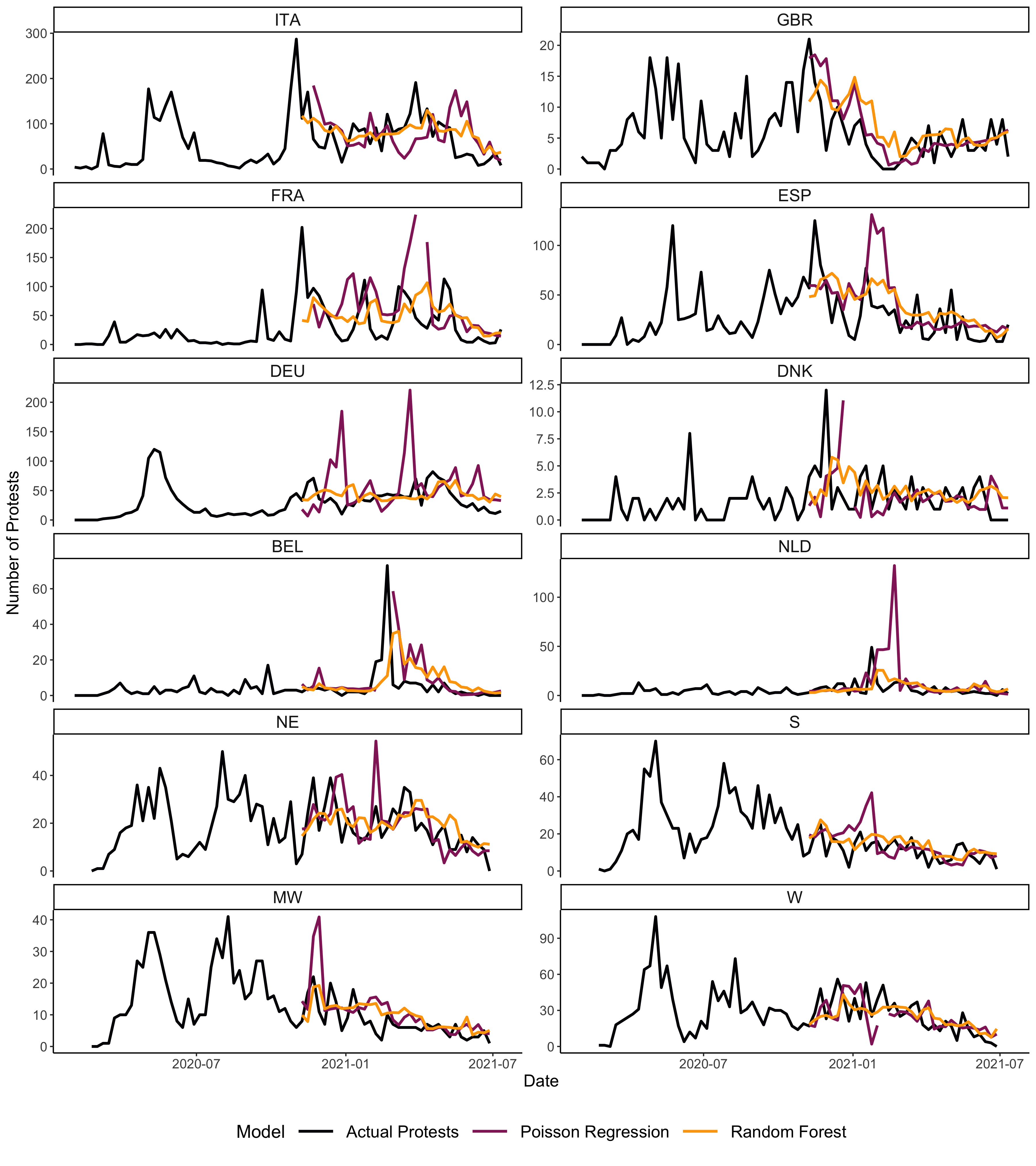}
\caption{{\bf Two week forecasts.}
This plot shows the two week protest count forecasts for the regions of interest. Color indicates forecast method. Points of severe overestimation were excluded for visualization purposes. This results in a break in the Poisson regression forecast for select regions (Italy, France, Denmark, and Belgium, and United States West).}
\label{fig7}
\end{figure}

\begin{figure}[!h]
\includegraphics[width=.85\textwidth, center]{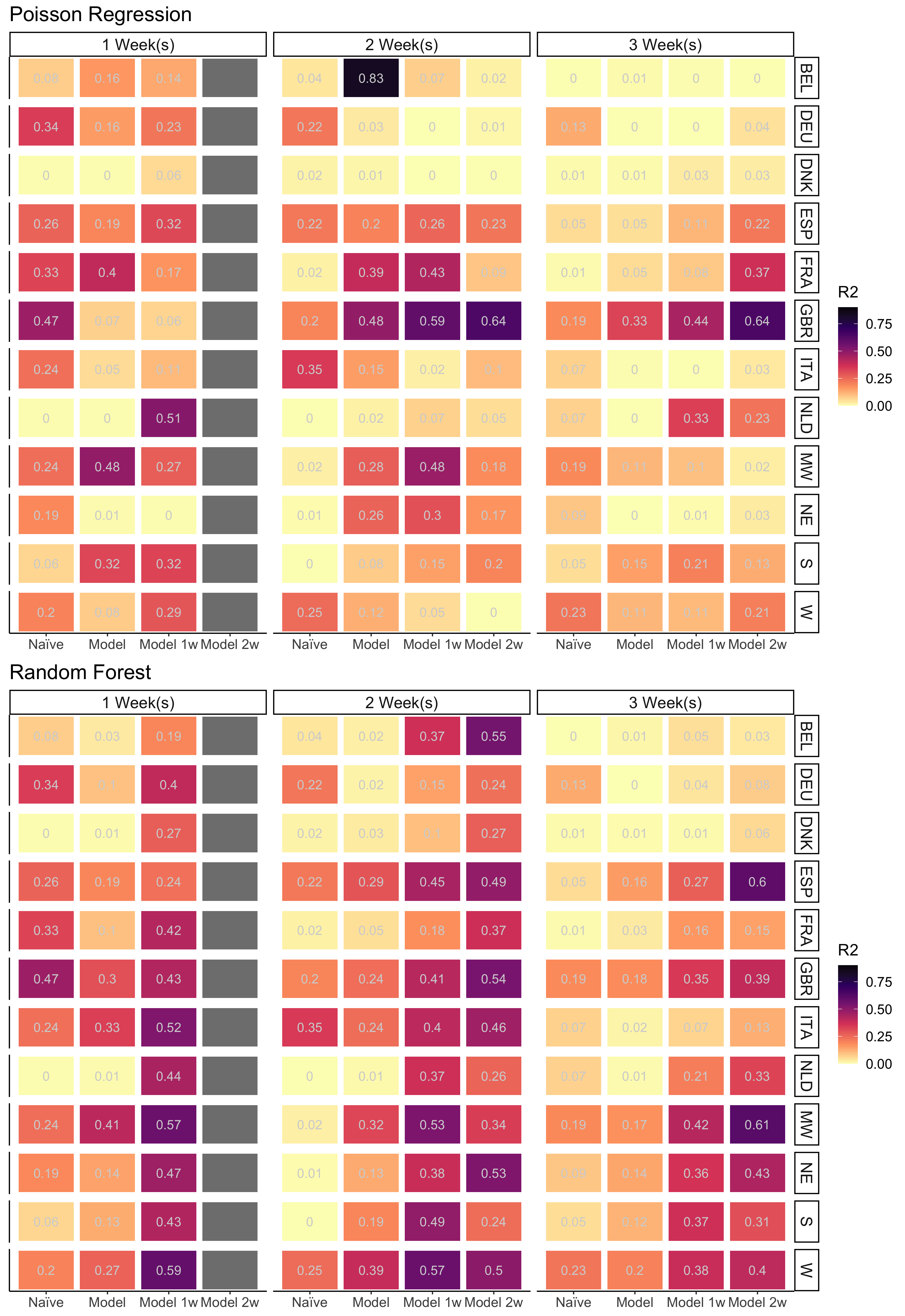}
\caption{{\bf Pearson's $R^2$ from protest count forecast.}
This plot shows the $R^2$ from forecasts of protest counts. The top plot depicts the results from the Poisson regression and the bottom plot depicts the results from the random forest. The horizontal axis labels the model method and model shifts (ex. Model 1W is the model forecast shifted one week back). Note that shifting the model also effectively alters the forecast period (ex. a one week forecast shifted back one week is a nowcast). Each colored box depicts the $R^2$ for a forecast method at a specific forecast time frame (column) and region (row). The boxes are colored and labeled with the $R^2$ (darker color: better $R^2$).}
\label{fig8}
\end{figure}

The forecasts of protest counts in Europe tended to have low $R^2$s (Fig \ref{fig7}, \ref{fig8}). The forecasts had larger $R^2$s when shifted back, effectively shortening the forecast period. For example, the $R^2$s for the one week forecast were highest when shifting the forecast back one week, effectively becoming a nowcast. Overall, the Poisson regression obtained the highest $R^2$s (e.g., $R^2 = 0.83$ for Belgium for the two-week forecast), but random forest obtained higher forecasting $R^2$s in general and across countries. Shifting the correlations back either one or two weeks improved $R^2$ values for all forecast models and times. Germany and Denmark were the countries with the lowest correlations across the forecasting models and time periods, however, the majority of forecasts did not achieve an $R^2 > 0.5$ across all countries.

In the U.S., the forecast of protest counts similarly had higher $R^2$s for the random forest than the Poisson regression. The Midwest exhibited the highest $R^2$s for the Poisson regression (e.g., $R^2 = 0.48$ for the one-week forecast), but all $R^2$s for the Poisson regression in U.S. regions were less than $0.5$. The Midwest also obtained the highest $R^2$ for the random forest (e.g., $R^2 = 0.61$ for the three-week forecast shifted back two weeks), although the West was not far behind with $R^2 = 0.59$ and $R^2 = 0.57$ for the one-week forecast and two-week forecast both shifted back one week. The random forest performed best for the Midwest and the West. The $R^2$ values for the South and the Northeast predominantly fell below $0.5$.

\begin{figure}[!h]
\includegraphics[width=.85\textwidth, center]{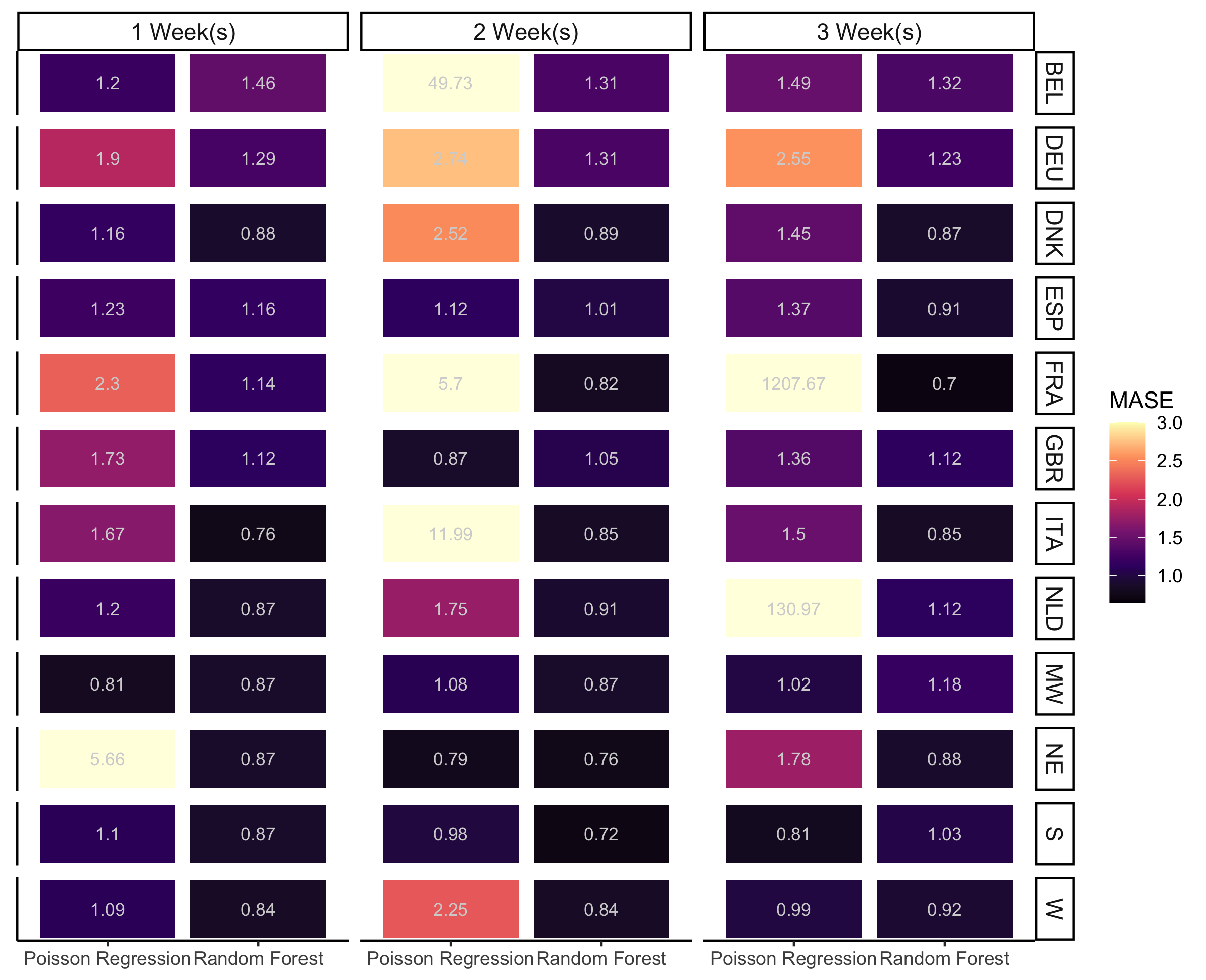}
\caption{{\bf Mean absolute scaled error from protest count forecast.}
This plot shows the mean absolute scaled accuracy (MASE) from forecasts of a binary protest response. The horizontal axis labels the model method. Each colored box depicts the MASE for a forecast method at a specific forecast time frame (column) and region (row). A value less than $1$ has a lower error than the corresponding naïve forecast. The boxes are colored and labeled with the MASE (darker color: better MASE).}
\label{fig9}
\end{figure}

The random forest forecasts did outperform naïve models in terms of mean absolute error (MAE) in certain countries, while the Poisson regression tended to overestimate the number of protests leading to high MAE, and therefore, high mean absolute scaled error (MASE) (Fig \ref{fig9}). Forecasts for Denmark and Italy had MASE $< 1$ for all random forest models, while France and the Netherlands had MASE $< 1$ for one- and two-week random forest forecasts. Spain's three-week random forest forecast had a MASE $< 1$, and Great Britain two-week Poisson forecast was the only Poisson forecast to obtain a MASE $< 1.$ The forecasts for Belgium and Germany never achieved a MASE $< 1$. 

United States regions had similar results, with the MASE for the Poisson regression tending to be higher than the MASE for the random forest. At the one- and two-week forecasts, the MASE $< 1$ for each region's random forest model. At the three-week forecast, the Northeast and West random forest models also achieved MASE $< 1$. The Midwest, Northeast, and South each had a single Poisson regression MASE substantially less than $1$, in the one-, two-, and three-week forecasts, respectively. Overall, the forecasts of protest counts were relatively successful based on MASE scores for most regions in Europe and the United States, however, $R^2$ forecast values were low across all regions.

\section*{Discussion}

We found that COVID-19 cases and deaths, public health orders, and misinformation contributed to COVID-19-related protests. Specifically, COVID-19-related protests tended to increase concurrently with COVID-19 cases towards the end of 2020 into 2021. This may indicate that many of the protests against strict public health policy at this time represented pandemic fatigue, rather than an objection that the policies were unnecessary, given the rising cases. In addition, some public health orders, such as stay-at-home and closures, were correlated with COVID-19-related protests. Other orders, such as vaccine and mask policies, resulted in an increase in protests directly after policy enactment or strengthening. COVID-19, lockdown, mask, and vaccine misinformation, measured with Google Trends data, were also highly correlated to COVID-19 protests, although these relationships differed by region. 

Despite the relationships seen in exploratory analysis, the predictive capacity of these data streams varied by region; some regions had forecast accuracy comparable to or higher than relevant literature, while others demonstrated almost no relationship to protests ~\cite{Bahrami2018, Kallus2014,Qiao2017}. Specifically, for a binary protest response both logistic regression and random forest models performed well for Belgium, Great Britain, Spain, and France. The forecasts for the other four European countries performed poorly and demonstrated no relationship with the binary protest outcome. For the United States, the South was the only region were the binary protest forecasting model performed well, and there it performed extremely well at two and three week forecasts. Forecasting protest counts had limited success in correctly predicting the pattern of protest counts as measured by $R^2$. While the Poisson regression forecasts did achieve some high $R^2$ values in European countries, this method tended to produce a more erratic forecast that tended to overestimate, causing very high mean absolute model error. The random forest model produced more consistent results: forecasts obtained $R^2$ around $0.5$, and forecasts had lower model error than the naïve model in around half of the European countries of interest for each forecast time period and all regions of the United states at one- and two-week forecasts. The forecasts for the United States generally were unable to predict times of high protest, but were successful at predicting protest counts, while forecast accuracy differed more by region than by protest outcome for the European countries. Although there was spatial heterogeneity in the forecasting ability of the data streams, in some regions, we were able outperform a naïve model in predicting both times of high protests and the overall trend in protest counts.

The variable success of the forecasts may be due to the specific difficulties of COVID-19 data and the limitations of the data sources. Due to the fast evolution of the COVID-19 pandemic, public health orders and misinformation that may be related to COVID-19 protest in one time period, may not even be relevant a month later. Therefore, features selected for the models in the training period may not be relevant in the testing period. While we used a rolling training window to mimic what would be done operationally, it is possible that our training window was too long to identify abrupt changes of interest in specific COVID-19 topics that had forecasting potential in the testing period. One example of this can be seen in Fig 2H: the Google Trends school search term feature is relatively unrelated to COVID-19-related protests in the Netherlands in 2020, however, it is highly related starting at the beginning of 2021. In addition, the three European countries that had the most significant predictors across the entire time period (Germany, Denmark, and the Netherlands), had some of the worst performing binary outcome forecasts. This fact is further indication that even data streams that were highly related with COVID-19 protest counts over the whole time frame did not necessarily correspond to successful forecasts. In addition, exploratory analysis demonstrated that while there was a time lag in the relationship between COVID-19 cases and public health orders in the beginning of 2020, there was almost no time lag in these relationships towards the end of 2020 into 2021. Therefore, the data may be more applicable for nowcasting, rather than forecasting. This is also reflected in the fact that the $R^2$s of the protest count forecasts tended to be higher when the the forecasts were shifted back a week or two, effectively shortening the forecast period.

Limitations of the data sources may have also influenced forecast success, with one of the largest limitations being the translation of the Google Trends search terms. While Google Translate has been shown to translate some short words and phrases correctly, it fails to translate idiomatic, contextually specific, and culturally relevant expressions ~\cite{Amilia2020}. While many search terms we selected are short words in phrases, phrases associated with COVID-19 are clearly associated within a unique cultural context and are specific to the COVID-19 policies and outcomes in a region. The fact that countries have created new words to describe aspects of the COVID-19 pandemic makes the issue of translation even more challenging ~\cite{oxfordlanguage}. Therefore, while significant efforts were made to translate phrases correctly, there still may be errors within the translation of search terms that could have negatively impacted the Google Trends data stream for non-English speaking countries. In fact, the forecasts for Great Britain, the only predominantly English speaking European country we examined, performed the best in the majority of the model evaluation metrics compared to the other European countries. The protest count forecasts for the United States also generally performed better than the forecasts for the European countries. We also chose to standardize the search terms across all regions, which may have excluded common phrases associated with COVID-19 misinformation specific to certain countries. In addition, the search terms were selected informally, and thus, they may not have fully encompassed the types of misinformation that we desired to capture in general. We also included terms not directly related to misinformation, with the assumption that increase in attention to a topic coincides with increase in misinformation about that topic, however this assumption needs to be explored further in a pandemic situation when attention is generally high. 

Furthermore, while Google Trends data has been used successfully to measure and forecast diseases and protests, the data itself has several limitations ~\cite{Verma2018, Timoneda2021}. Google Trends data is only suggestive of overall trends in attention to misinformation. The time series trends are dependent on the dates over which the data is pulled, and the exact process that Google uses to normalize the data is unknown. In addition, the data is dependent on the assumption of widespread internet use and the use of Google as a main search engine. Furthermore, the OxCGRT data reported when a stronger/weaker policy was \textit{enacted}, not when it was \textit{announced} which may have limited the data source's potential for forecasting: many individuals react to policy changes on announcement, which could be weeks ahead of when it is enacted. We also made the conservative assessment of filling missing policy data with zeros, however this is most likely not an accurate representation of policy strength at those times. Finally, the COVID-19 case data was not adjusted for the different testing strategies in individual regions. Therefore reported COVID-19 cases may have a different meaning between our various regions of interest.

We suggest that future work examine new methods of adapting model fitting when modeling situations that evolve quickly. For example, implementing unsupervised learning to select social media data, instead of identifying specific phrases, could help capture the nuance of discussions/topics that evolve over time on social media. This could resolve the issue of selecting model features that are not relevant to the model test period. In addition, processing social media data in the native language could help alleviate the errors that may be created in both the selection of search terms and translation. Our analysis has demonstrated that there exist complex and intertwined relationships between health outcomes, governmental response, and societal reaction during times of significant change, such as the COVID-19 pandemic, and that the exploration of these relationships in future contexts requires further methods.

\section*{Conclusion}
We found that COVID-19 health outcomes, public health orders, and misinformation contributed to COVID-19-related protests, however, these relationships evolved throughout time as the pandemic progressed. Specifically, closure policies and misinformation tended to be the most related to protests across all regions of interest. These data streams were successful at forecasting COVID-19-related protests as both a binary response and protest count for up to three weeks for several regions. However, forecast accuracy varied substantially by region, and we believe this may be due to translation issues within our misinformation data as well as the difficulty of selecting the best model features in evolving situations. Diverse data streams, including social media data, have the capacity to forecast protests, and are integral to capturing a rapidly changing situation with impacts that extend across various sectors of society. 

\section*{Data Availability Statement}
All data used in this analysis are publicly available. Details are provided in \nameref{S1_Table1} Table 1.

\section*{Supporting information}

\begin{table}[!ht]
\label{S1_Table1}
{\bf SI Table 1. Data Specifics.}{ All data was collected at the national level for Europe and state level for the U.S.}
\begin{adjustwidth}{-0.75in}{0in}
\centering
\begin{tabular}{|l|l|l|l|l|l|l|}
\hline
{\bf Data Source} & \bf{Data Details} & {\bf Country} & {\bf Start} & {\bf End} & {\bf Resolution }& {\bf Source}\\ 
\thickhline
\multirow{1}{4em}{ACLED} & Acts of social unrest & Belgium &  2020-01-01 & 2021-07-30 & Daily & ~\cite{acled2021}\\ 
&  & Germany &  2020-01-01 & 2021-07-30 & & \\ 
&  & Denmark &  2020-01-01 & 2021-07-30 & & \\ 
&  & Spain &  2020-01-01 & 2021-07-30 & & \\ 
&  & France &  2020-01-01 & 2021-07-30 & & \\ 
&  & Great Britain &  2020-01-01 & 2021-07-30 & & \\ 
&  & Italy &  2020-01-01 & 2021-07-30 & & \\ 
&  & Netherlands &  2020-01-01 & 2021-07-30 & & \\ 
&  & U.S. MW &  2020-01-01 & 2021-06-25 & & \\ 
&  & U.S. NE &  2020-01-01 & 2021-06-25 & & \\ 
&  & U.S. S &  2020-01-01 & 2021-06-25 & & \\ 
&  & U.S. W &  2020-01-01 & 2021-06-25 & & \\ \hline
\multirow{1}{4em}{OxCGRT} & Public health order strength & Belgium &  2020-01-01 & 2021-08-16 & Daily & ~\cite{oxcgrt_git2020, jh2021}\\ 
& (see Table \ref{table2}). Also includes & Germany &  2020-01-01 & 2021-08-16 & & \\ 
& Johns Hopkins COVID-19& Denmark &  2020-01-01 & 2021-08-16 & & \\ 
& case and death data. & Spain &  2020-01-01 & 2021-08-16& & \\ 
&  & France &  2020-01-01 & 2021-08-16 & & \\ 
&  & Great Britain &  2020-01-01 & 2021-08-16 & & \\ 
&  & Italy &  2020-01-01 & 2021-08-16 & & \\ 
&  & Netherlands &  2020-01-01 & 2021-08-16 & & \\ 
&  & U.S. MW &  2020-01-01 & 2021-08-16 & & \\ 
&  & U.S. NE &  2020-01-01 & 2021-08-16 & & \\ 
&  & U.S. S &  2020-01-01 & 2021-08-16 & & \\ 
&  & U.S. W &  2020-01-01 & 2021-08-16 & & \\ \hline
\multirow{1}{4em}{Google Trends} & See \nameref{S1_Table2} Table 2 for & Belgium &  2020-01-05 & 2021-08-01 & Weekly & ~\cite{gtrends}\\ 
& specific search terms. & Germany &  2020-01-05 & 2021-08-01 & & \\ 
& & Denmark &  2020-01-05 & 2021-08-01 & & \\ 
& & Spain &  2020-01-05 & 2021-08-01& & \\ 
&  & France &  2020-01-05 & 2021-08-01 & & \\ 
&  & Great Britain &  2020-01-05 & 2021-08-01 & & \\ 
&  & Italy &  2020-01-05 & 2021-08-01 & & \\ 
&  & Netherlands &  2020-01-05 & 2021-08-01 & & \\ 
&  & U.S. MW &  2020-01-05 & 2021-08-01 & & \\ 
&  & U.S. NE &  2020-01-05 & 2021-08-01 & & \\ 
&  & U.S. S &  2020-01-05 & 2021-08-01 & & \\ 
&  & U.S. W &  2020-01-05 & 2021-08-01 & & \\ \hline
\end{tabular}
\end{adjustwidth}
\end{table}

\begin{table}[!ht]
\label{S1_Table2}
{\bf SI Table 2. Google Trends Search Terms.}
\begin{adjustwidth}{-0.75in}{0in}
\centering
\begin{tabular}{|l|l|}
\hline
{\bf Category} & {\bf Search Words}\\ 
\thickhline
\multirow{2}{4em}{General} & 'fake news', 'false information', 'government lies', 'scaremongering',  'conspiracy theory',\\ & 'government trust', 'misinformation', 'media lie', 'fake news media', 'media trust' \\ \hline
\multirow{4}{4em}{COVID-19} & 'covid lie', 'covid idiot', 'covid denier', 'covid skeptic', 'covid conspiracy theory', 'covid conspiracy', \\& 'covid hoax', 'covid is a hoax', 'covid is a lie', 'covid fake', 'fake covid', 'coronavirus lie','coronavirus idiot', \\& 'coronavirus denier', 'coronavirus skeptic', 'coronavirus conspiracy theory', 'coronavirus conspiracy',\\& 'coronavirus hoax', 'coronavirus is a hoax', 'coronavirus is a lie', 'coronavirus fake', 'fake coronavirus' \\ \hline
\multirow{2}{4em}{Lockdown} & 'lockdown too long', 'anti lockdown', 'lockdown', 'stay at home', 'restrict freedom', 'coronavirus anti lockdown', \\& 'covid anti lockdown', 'coronavirus lockdown', 'covid lockdown', 'lockdown too strict', 'when is lockdown over' \\ \hline 
\multirow{3}{4em}{School} & 'schools should be open', 'school closure', 'covid school closure', 'coronavirus school closure', 'kids coronavirus', \\& 'kids covid', 'schools open',  'coronavirus schools open','covid schools open', 'kids need school', \\& 'coronavirus hoax school', 'covid hoax school', 'open school' \\ \hline
\multirow{3}{4em}{Mask} & 'anti mask', 'mask cant breathe', 'no mask mandate', 'mask', 'masks', 'face mask', 'face masks', 'mask freedom',  \\& 'mask lie', 'mask conspiracy', 'mask mandate', 'anti masks', 'anti face mask', 'mask uncomfortable',  \\& 'mask exemption', 'mask hoax' \\ \hline
\multirow{5}{4em}{Vaccine} & 'covid vaccine not safe', 'vaccine hoax', 'vaccine lie', 'covid vaccine', 'coronavirus vaccine', 'vaccine chip', \\& 'coronavirus vaccine lie', 'covid vaccine lie', 'pfizer', 'pfizer vaccine', 'moderna', 'moderna vaccine', \\& 'johnson \& johnson', 'astrazeneca', 'astrazeneca vaccine', 'johnson \& johnson vaccine', 'johnson and johnson', \\& 'johnson and johnson vaccine','astrazeneca not safe', 'coronavirus vaccine death', 'covid vaccine death', \\& 'johnson and johnson not safe', 'covid vaccine conspiracy', 'coronavirus vaccine conspiracy' \\ \hline
\multirow{3}{4em}{Economic} & 'covid help', 'corona help', 'income support', 'apply for unemployment', 'debt relief', 'unemployment', \\& 'open economy' 'coronavirus help', 'open store', 'open restaurant', 'need to open economy',  \\& 'open stores', 'open restaurants' \\ \hline
\end{tabular}
\end{adjustwidth}
\end{table}

\begin{table}[!ht]
\begin{adjustwidth}{-0.75in}{0in}
\label{S1_Table3}
{\bf SI Table 3. Full set of possible predictors.}
\centering
\begin{tabular}{|l|l|}
\hline
{\bf Data Stream} & {\bf Predictors}\\ 
\thickhline
COVID-19 Health Outcomes & weekly cases, weekly deaths \\ \hline
\multirow{7}{4em}{Public Health Orders} &  C1 School closing, C2 Workplace closing, C3 Cancel public events, \\& C4 Restrictions on gatherings, C5 Close public transport, C6 Stay at home requirements, \\& C7 Restrictions on internal movement,  C8 International travel controls, E1 Income support, \\& E2 Debt/contract relief,  H1 Public information campaigns, H2 Testing policy, \\&  H3 Contact tracing, H6 Facial Coverings, H7 Vaccination policy, \\& H8 Protection of elderly people, Stringency Index, Government Response Index, \\&  Containment Health Index, Economic Support Index \\ \hline
\multirow{2}{4em}{Google Trends} & general terms, COVID-19 terms, lockdown terms, school terms, mask terms,  \\& vaccine terms, economic terms \\ \hline 
\end{tabular}
\end{adjustwidth}
\end{table}

\section*{Acknowledgments}
The authors acknowledge support from the Laboratory Directed Research and Development (LDRD) program under a Director’s Initiative Project 20210766DI. This work is approved for distribution under LA-UR-21-29745. The findings and conclusions in this report are those of the authors and do not necessarily represent the official position of Los Alamos National Laboratory. Los Alamos National Laboratory, an affirmative action/equal opportunity employer, is managed by Triad National Security, LLC, for the National Nuclear Security Administration of the U.S. Department of Energy under contract 89233218CNA000001. The authors would like to thank Dr. Lauren Beesley VanDervort for her expertise and guidance in statistical methods.

\nolinenumbers

\end{document}